\documentclass[5p,times]{elsarticle}
\usepackage{lineno,hyperref}
\modulolinenumbers[5]
\journal{Journal of Magnetism and Magnetic Materials}

\begin{document}

\begin{frontmatter}

\title{\textit{Influence of the different strains' components on the uniaxial magnetic anisotropy parameters for a (Ga,Mn)As bulk system: a First-Principles Study.}}

\author[aa]{M.~Birowska\corref{cor1}}
\ead{Magdalena.Birowska@fuw.edu.pl}
\address[aa]{Institute of Theoretical Physics, Faculty of Physics, University of Warsaw, ul. Pasteura 5, PL-02-093 Warsaw, Poland}

\begin{abstract}
We present a computational study of the magnetic anisotropy energy for a given concentration of the Mn-ions in the GaAs host, in the framework of the density functional theory. We focus on the influence of a different kind of strains: biaxial, shear, and hydrostatic on the uniaxial magnetic anisotropy parameters $K_1$ and $K_2$, which reflect the magnetic anisotropy energy out- and in- (001) plane, respectively. We have shown that the general trends for the applied biaxial strain on anisotropy parameters are consistent with the experimental data. We have predicted the critical strains, for which the magnetization vector changes its direction. Our results have shown that it is not possible to modify considerably the uniaxial magnetic anisotropy parameters, exposing (Ga,Mn)As to hydrostatic pressure of a magnitude reasonable from experimental point of view.
\end{abstract}


\end{frontmatter}

\section{Introduction}
(Ga,Mn)As is the most intensively studied material among the III-V Dilute Magnetic
Semiconductors (DMS), and is considered as theirs flagship example \cite{RevModPhys.78.809,RevModPhys.86.187,RevModPhys.86.855}. Despite the fact that (Ga,Mn)As exibits too low Curie temperature $T_C$ for widespread applications, with the record around 200 K \cite{NanoLett.11.2584}, it remains the best test-bed material for new concepts in spintronics \cite{RevModPhys.86.855}. The spin-orbit mediated coupling of magnetic and semiconductor properties for this material gives rise to a large number of phenomena with a vast scope of possible applications \cite{Nature.455.2008}. One of the intriguing phenomena in this material is the presence of uniaxial magnetic anisotropy in structurally nominally cubic dilute magnetic semiconductor \cite{PhysRevLett.90.107201,PhysRevLett.90.167206,PhysRevB.71.121302,PhysRevB.67.205204,PhysRevLett.97.077201}. Uniaxial magnetic anisotropies in (Ga,Mn)As samples have been proved by numerous experiments, and by many different characterization methods, among others magnetotransport \cite{PhysRevLett.90.107201,1367-2630-10-5-055007,PhysRevB.79.195206}, magneto-optics \cite{PhysRevLett.90.167206,aip.81}, magnetic measurements using the superconducting quantum interference device (SQUID) \cite{PhysRevB.71.121302}, as well as ferromagnetic resonance (FMR) \cite{PhysRevB.67.205204}. 

Moreover, it is experimentally well-known that magnetic anisotropy in (Ga,Mn)As, depends on the substrate lattice constants, temperature and hole concentration \cite{PhysRevLett.95.217204,PhysRevB.70.245325,PhysRevLett.95.187206}. Moreover, it has been found that magnetic anisotropy can be controlled by the epitaxial growth. Generally, (Ga,Mn)As samples grown on the (In,Ga)As substrate are tensily strained, and the easy axis of magnetization points perpendicular to the (001) plane \cite{PhysRevB.71.241307,jap.98.1063}; whereas samples grown on the GaAs substrate are compressively strained, and the easy axis of magnetization lies in the (001) plane \cite{PhysRevB.70.245325,PhysRevLett.93.117203}. 

The understanding of the underlying mechanism of the uniaxial magnetic anisotropies in this prototypical DMS is of crucial importance for the potential applications in magnetic recording technologies. Therefore, in this paper, we focus on the impact of the different external strains on the uniaxial magnetic anisotropy parameters $K_1$ and $K_2$. We do not consider here the problem of the origin of the uniaxial magnetic anisotropy in (Ga,Mn)As. This problem is widely discussed elsewhere \cite{Kong.97.2005,Piano.98.2011,Moosbuhler.91.2002}. In the paper \cite{PhysRevLett.108.237203} the authors showed that non-random Mn-ions distribution that sets-in at the growth surface during epitaxy is the physical origin of the bulk uniaxial magnetic anisotropy in-plane and out-of-plane. It is worth to mention that although, the existence of the surface guarantees the symmetry low enough to induce uniaxial magnetic anisotropy, the experiments unambiguously clarify that this phenomenon has to be ascribed to bulk but
not surface or interface property of a (Ga,Mn)As \cite{PhysRevLett.90.167206, PhysRevB.71.121302, PhysRevB.80.155203}. Building on these assumption we calculate magnetic anisotropy energy (MAE) of a bulk crystal for fixed non-random concentrations of Mn-ions in the framework of Density Funcional Theory (DFT) \cite{PhysRev.136.B864,PhysRev.140.A1133}, differently then it was presented previously in the paper \cite{PhysRevLett.108.237203}.

The paper is organised as follows. First, in this letter, we provide a theoretical description of the self-consistent method to calculate magnetic anisotropy energy. Then we define the anisotropy parameters and strain tensor. Furthermore, we show how the uniaxial magnetic anisotropy parameters can be modified by external factors like strains by means of the ab initio method. Finally we compare our results with experiment. We want to stress here that to the best of our knowledge, the present letter is the first  ab initio study of the impact of different strains' components on the magnetic anisotropy energy up to date. 

\subsection{Self-consistent method to calculate magnetic anisotropy energy}

\begin{figure} 
\centering
\includegraphics[width=8.6 cm]{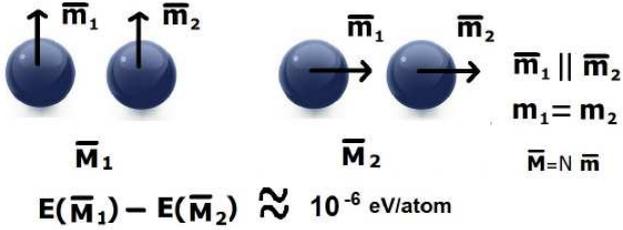}
\caption{\label{MAE} Only collinear alignment of magnetization is assumed for our system $\vec{m}{_1}$ is parallel to $\vec{m}{_2}$. The magnetic anisotropy energy is the work needed to rotate a magnetization vector from one direction (here $\vec{M}{_1}$) to another ($\vec{M}{_2}$). The black arrows on the balls indicate the magnetic moments of the Mn-ions.}
\end{figure}

In order to calculate magnetic anisotropy energy\footnote{In our calculations magnetic anisotropy refers to magnetocrystalline anisotropy.} (MAE), we have generated fully relativistic pseudopotentials to account for the spin-orbit interaction and employ the local density approximation (L(S)DA) for exchange-correlation density functional with the effects of non-collinear magnetism included \cite{PhysRevB.71.115106}. All of our MAE results are obtained employing the planewave code, Quantum Espresso (QE) \cite{QE}. We have performed the calculations for collinear alignment of two magnetization vectors coming from the Mn-ions, as it schematically presented in the Figure~\ref{MAE}. During the self-consistent loop total magnetization direction evolves by redirecting its magnetization to be parallel to an easy magnetization axis (ground state). 
In  order to calculate MAE it is necessary to obtain the total energy
for the magnetization direction different from the ground state, namely parallel to the hard axis of 
total magnetization. The way to do so is to have the initial geometry of magnetization the same like the final one, in other words to constraint the magnetization direction. However, we notice that we need a large number of iteration steps to reach 
the ground state of magnetization direction, and it is actually impossible to reach by using supercomputers provided today. The self-consistent method converges very slowly to the ground state configuration.
In addition, the total magnetization direction, and hence the total energy, changes insignificantly at each step. Therefore, it is not necessary to constrain the total magnetization direction. It is sufficient to choose the initial magnetization and energy convergence threshold small enough to predict accurately the absolute value of the atomic magnetic moments and their direction, and simultaneously sufficiently large not to change total magnetization direction from the initial condition.
Therefore, we rotate magnetization vector, coming from the \textit{3d} states of the Mn-ions, relative to the crystallographic
axes as it is shown in Figure~\ref{magnetyzacja}, but, for each direction, the calculations have been performed
self-consistently. We noticed that for different direction, 
the optimized magnitude of magnetization on each of the Mn-atoms is the same within the error of 0.0001 $\mu_B$/atom which changes the energy by up to 0.2 $\mu$eV/cell. Those are two orders of magnitudes smaller than we need. We have carefully checked the convergence of the MAE versus the parameters which could influence our results.

Our procedure is to determine the anisotropy parameters for 6.25 \% concentration of Mn-atoms in the cubic supercell\footnote{This is a common concentration of Mn-atoms in a real (Ga,Mn)As samples.}, which contains 64 atoms. Mn-pair is placed at the nearest neighbors (NN) position, which is the most energetically stable configuration from all non-equivalent ones for a given supercell. For the NN configuration, there are two non-equivalent crystallographic directions: [110] and [1$\bar{1}$0]. Our previously reported results \cite{PhysRevLett.108.237203} show that the Mn-pairs choose preferable positions already during the growth process, just approaching the surface of the semiconductor, which results in a non-random Mn distribution at the growth surface during epitaxy. Therefore, the Mn-pair was aligned along [1$\bar{1}$0] crystallographic direction. We determine a sufficient kinetic energy cutoff of 40 Ry for the plane-wave expansion of the pseudo-wave function and 160 Ry for the charge densities. For the k-points summation, we use $14\times14\times14$ Monkhorst-Pack mesh \cite{PhysRevB.13.5188}. Smearing parameter of 0.0001 Ry with Fermi-Dirac function \cite{PhysRevB.40.3616} is used for calculation of the electron density in the case of metallic systems, where efficient dealing with the Fermi surface is necessary.

We also want to note here that the spin magnetic moments are always predicted, but there are no orbital magnetic moments taken into account. The orbital magnetic moments can be experimentally measured, and it might be an interesting issue to compare it with theoretically predicted values. Furthermore, its influence on the MAE is worth a study.

\subsection{Anisotropy parameters}

\begin{figure} 
\centering
\includegraphics[width=0.3\textwidth]{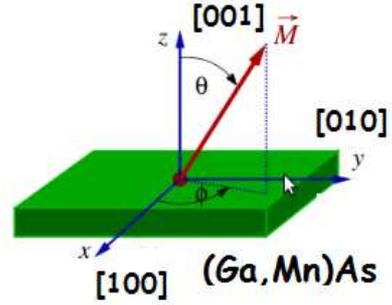}
\caption{\label{magnetyzacja} Geometry for calculating magnetocrystalline anisotropy. The Cartesian coordinate system chosen to derive the expression for the free energy for different crystal symmetries, where $\vartheta$ is a polar angle and $\varphi$ azimuthal angle.}
\end{figure}

Phenomenologically, the crystallographic easy axis of the magnetization is determined by the minimum of free energy $F(\overrightarrow{\Omega_M})$  as a function of direction of magnetization  \\
$\overrightarrow{\Omega_M}=(\sin(\vartheta)\cos(\varphi),\sin(\vartheta)\sin(\varphi),\cos(\vartheta))$, and can be expanded into invariants $f_i$:
\begin{equation}
\label{basis}
 F_{(\vartheta,\varphi)}=\sum_{i}K_{i}f_{i}
\end{equation} where $f_i$ are the basis invariants which are defined in respect to decomposition of the space of spherical harmonics with given $l$ into irreducible representations of the $T_d$ group and $K_i$ are anisotropy parameters, as it was previously considered in the paper \cite{PhysRevLett.108.237203}. The magnetization vector is presented in the Figure~\ref{magnetyzacja}. 

\begin{figure} 
\centering
\includegraphics[width=0.55\textwidth]{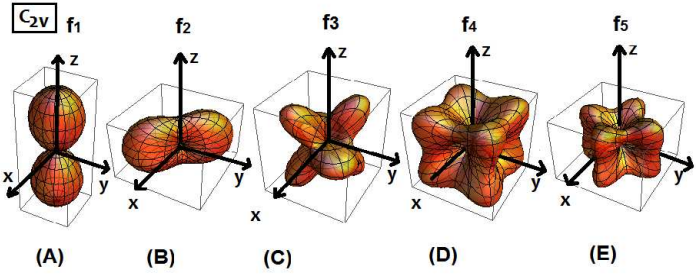}
\caption{\label{orders} Energy surfaces for a $C_{2v}$ symmetry. (A) Uniaxial out-of-plane anisotropy, (B) uniaxial in-plane anisotropy, (C) fourth-order correction to the uniaxial in-plane anisotropy, (D) cubic anisotropy, and (E) fourth-order correction to the uniaxial out-of-plane anisotropy.}
\end{figure}

 The Mn-pair at the NN position, which we consider in this paper, exhibits the $C_{2v}$ point group symmetry, and the basis invariants  up to the fourth-order $(l=4)$ are presented in the Figure~\ref{orders}. The $f_{0}$ is a parameter, $f_{1}$ is the uniaxial [001] out-of (001) plane anisotropy with its correction $f_{5}$ ($4^{th}$ order). The $f_{2}$ is the in-plane uniaxial [110] anisotropy, and its correction is $f_{3}$ ($4^{th}$ order term). $f_{1}$ and $f_{2}$ are the second-order terms, and $f_4$ is a cubic anisotropy term as it is widely explained in the paper \cite{PhysRevLett.108.237203}.

Our procedure is as following. First, we calculate the total energy for a different orientation of the magnetization vector (as a function of the angles $\vartheta$, $\varphi$). Then we fit our total energy results to the Eq. (\ref{basis}) expanded up to fourth-order terms for the  $C_{2v}$ point group symmetry. Then, from the fitting we obtain the anisotropy parameters.

\subsection{Strain tensor}

 The strain effects in films can be taken into account through the strain tensor $\hat{\varepsilon}$ \cite{Skierkowski.112.2007}:
\[\hat{\varepsilon}= 
 \left( \begin{array}{ccc}
  \varepsilon_{xx}  & \varepsilon_{xy} & \varepsilon_{xz} \\ 
 \varepsilon_{yx} & \varepsilon_{yy} &\varepsilon_{yz}\\ 
 \varepsilon_{zx}& \varepsilon_{zy} & \varepsilon_{zz}  \end{array} \right).\]

The lattice ${a_i}'$ and basis vectors $\tau_{j}'$ of a strained crystal \cite{Bir} are defined respectively:
\begin{equation}
{\overrightarrow{a_i}}'=(1+\hat{\varepsilon })*\overrightarrow{a_i},\\
\end{equation}
\begin{equation}
{\overrightarrow{\tau_{j}}}'=(1+\hat{\varepsilon })*\overrightarrow{\tau_{j}}+\overrightarrow{\eta_j },
\end{equation}
where  $\overrightarrow{\eta_j }$ is so-called internal strain parameter\footnote{In our calculations no internal strains have been taken into account.}, i.e., the shift of the atoms along the line that remains invariant under symmetry operations of the point group of the strained crystal, and ${\overrightarrow{a_i}}$ and ${\overrightarrow{\tau_i}}$ are lattice and basis vectors, respectively, of the unstrained lattice. The volume ${\Omega_{0}}'$ of the cell spanned on the lattice vectors  ${\overrightarrow{a_i}}'$ is expressed by the strain tensor:
\begin{equation}
{\Omega_{0}}'=\Omega_{0}(1+\mathrm{Tr}\hat{\varepsilon}),
\end{equation}
where $\Omega$ is the unit cell volume of the unstrained crystal. 
The symmetry of the structure defines the form of the strain tensor.

\section{RESULTS}

 This section is organized as follows. First, we consider Mn-pair without taking into account any strain components. Then various lattice deformations under biaxial, shear and hydrostatic strains are considered. We focus on their impact on the uniaxial magnetic anisotropy parameters $K_1$ and $K_2$. Then, we compare our results with the experiments taken from literature.

\subsection{Mn-pair in the simple cubic supercell with no external strains}
 \begin{figure}
\includegraphics[width=0.5\textwidth]{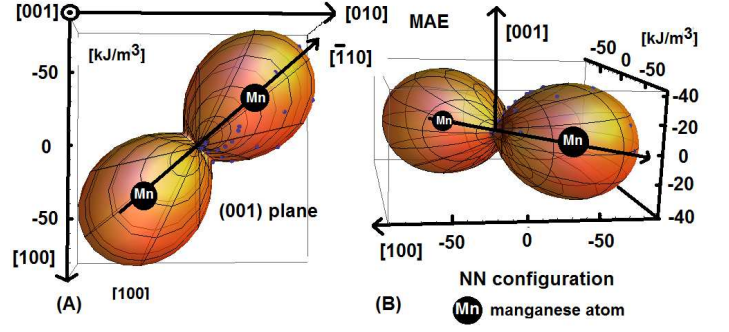}
\caption{\label{NNnew}Magnetic anisotropy energy surface of the nearest neighbor (NN) configuration of the Mn-pair at the [1$\bar{1}$0] crystallographic direction, as function of the spherical angles ($\theta,\phi$). The black balls indicate the positions of the Mn-atoms in the supercell. The magnetic ground state is for the Mn magnetic moments pointing along the [1$\bar{1}$0] direction, which is simultaneously the easy axis of the magnetization, whereas the [001] direction is the hard axis of magnetization.}
\end{figure}

Our results show that for the substitutional position of the NN Mn-pair, the system acquires maximal energy when magnetization is oriented along [001] direction (the hard axis of magnetization), whereas the most favorable direction of magnetization energetically (i.e., the easy axis) is along [1$\bar{1}$0] one, i.e, when magnetization is parallel to the line connecting the Mn-atoms within the pair, as it is presented in the Figure 4. The anisotropy parameters out-of-plane and in-plane are $K_1$=60.54 kJ/m$^3$, $K_2$=91.25 kJ/m$^3$, respectively. It is worth to mention that when the positions of the atoms are not relaxed, then the anisotropy parameters are smaller than the relaxed case, and equal to $K_1$=49.02 kJ/m$^3$, $K_2$=83.85 kJ/m$^3$. One can see that relaxation of the atoms increases both uniaxial terms $K_1$ and $K_2$, mostly due to decreasing distance between Mn-atoms by 0.1 \AA{}. The Mn-atoms during the optimization procedure move along [1$\bar{1}$0] crystallographic directions. Moreover, the As atom in between the Mn-atoms is displaced by 0.016 \AA{} along the [001] direction. One can notice that those displacements do not change the symmetry of the (Ga,Mn)As. The [001], [110] and [1$\bar{1}$0] are invariant directions for $C_{2v}$ symmetry. In other words, one can say that the optimization of the positions of atoms further enhances the magnetic anisotropy energy.

\subsection{Mn-pair with biaxial strain}
 
  \begin{figure}
  \centering
  \includegraphics[width=0.5\textwidth]{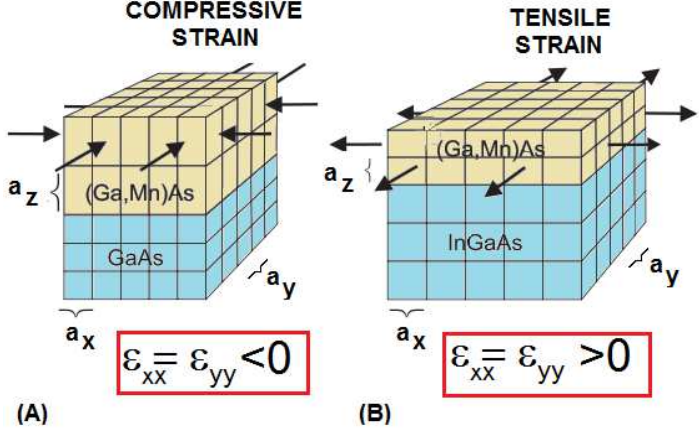}
  \caption{\label{CELLstrain} Schematic picture of the biaxial strain for the (Ga,Mn)As lattice. The arrows indicate the strain direction. (A) (Ga,Mn)As grown on a GaAs(001) substrate is under compressive strain. (B) (Ga,Mn)As grown on a (Ga,In)As(001) substrate is under tensile strain.}
\end{figure}

X-ray diffraction measurements showed that the out-of-plane lattice constants of (Ga,Mn)As films depend sensitively on the lattice constants of the substrate \cite{Welp.85.2004}. 

Due to the epitaxial growth of the (Ga,Mn)As films, at relatively low temperatures, the in-plane lattice constant is locked to that of the substrate. Therefore, (Ga,Mn)As film can be under tensile (grown on the (Ga,In)As) or compressive strain (grown on the GaAs), as it is presented in the Figure~\ref{CELLstrain}. Therefore, in this section we consider various lattice deformations under biaxial strains $e_{xx}$ (reflect the different substrates of (Ga,Mn)As films) in order to examine theirs influence on the magnetic properties of (Ga,Mn)As films. The biaxial strains can be considered as:
\begin{equation}
\label{exx}
\varepsilon_{xx}=\varepsilon_{yy} = \frac{a_0-a_{rel}}{a_0}, \\
\varepsilon_{zz}=-2\cdot\frac{C_{11}}{C_{12}}\cdot\varepsilon_{xx}, \\
\end{equation}
where $a_0$ and $a_{rel}$ are the lattice constants of the GaAs and (Ga,Mn)As film, respectively. $C_{11}$, $C_{12}$ are elastic stiffness constants with $\frac{C_{11}}{C_{12}}=0.453$, and here we assume that they are equal to those of GaAs. For example, biaxial strain components $\varepsilon_{xx}=\varepsilon_{yy}$ and $\varepsilon_{zz}\neq\varepsilon_{xx}$ appear when $T_{d}$ $\rightarrow$ $D_{2d}$ reduction of the symmetry takes place, whereas the shear component $\varepsilon_{xy}$ is a consequence of the $T_d$ $\rightarrow$ $C_{2v}$ symmetry reduction \cite{Skierkowski.112.2007}.

\begin{figure}
\centering
\includegraphics[width=0.5\textwidth]{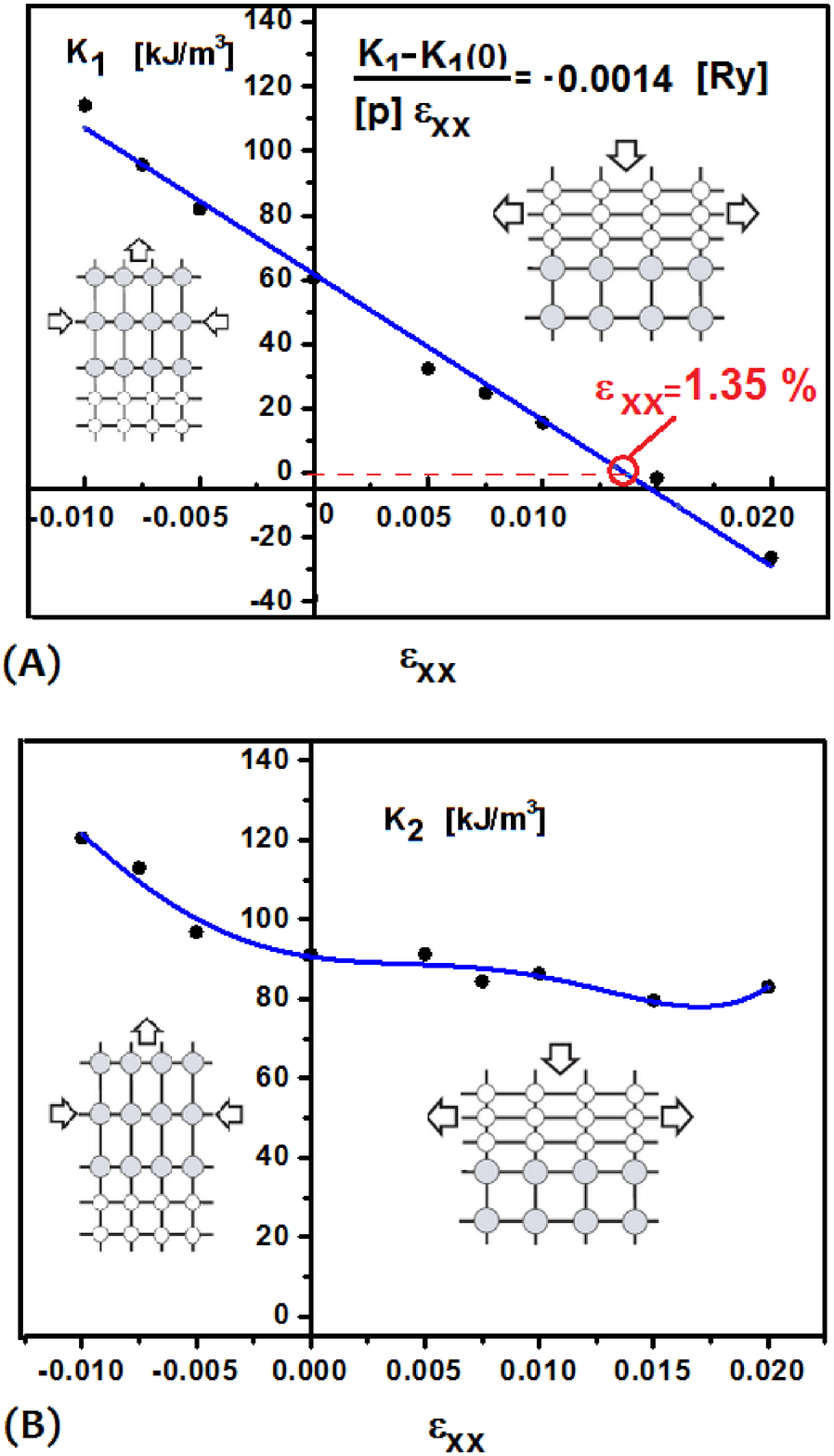}
\caption{\label{exxnew} The influence of the biaxial strain on the uniaxial anisotropy parameters. (A) The uniaxial out-of-plane anisotropy parameter $K_1$ as a function of biaxial strain for a Mn-concentration of $x$ = 6.25$\%$ (hole concentration is equal to $14.8\cdot10^{20}$ $cm^{-3}$). Linear dependency is clearly visible for a wide range of deformation strains and predicted slope is equal to $-0.0014$ Ry. For the increasing biaxial strain, the $K_1$ parameter decreases. For a critical value of $\varepsilon_{xx}$ = 1.35$\%$, the anisotropy parameter $K_1$ changes the sign, which indicates that out-of-plane magnetization direction is more energetically preferable in comparison to the in-plane [100] high symmetry magnetization direction. \textbf{(B)} - There is very weak nonlinear dependency of an in-plane uniaxial magnetic anisotropy $K_2$ on the biaxial strain (also mentioned in Ref. \cite{PhysRevB.79.195206}), generally decreasing with the increasing strain. Those results indicate that there is no possibility to change the sign of in-plane uniaxial parameter $K_2$ by the reasonable (experimentally accessible) range of biaxial strains.}
\end{figure}

Our results show a clear linear correlation of out-of-plane anisotropy parameter $K_2$ and biaxial strain, presented in Figure~\ref{exxnew} (A). According to the mean-field theory, the out-of-plane anisotropy parameter $K_1$ depends linearly on the hole concentration [p] and biaxial strain $\varepsilon_{xx}$ \cite{PhysRevB.79.195206,PhysRevB.63.054418}:
\begin{equation}
 K_1 = E \cdot e_{xx}\cdot [p].
\end{equation}

The authors \cite{PhysRevB.63.054418} predicted the value $E = -0.36$ Ry for the low concentration limit [p] = $10^{19}$ $cm^{-3}$, where only one hole subband was populated, whereas the Khazen \textsl{et al.} \cite{PhysRevB.77.165204} experimentally observed such linear behavior also for the high hole concentration (up to the $10^{21}$ $cm^{-3}$), and obtained the value $E = 0.0019$\footnote{In the paper \cite{PhysRevB.77.165204} the out-of-plane anisotropy parameter $K_1$ is indicated as $K_2\perp$.} Ry for the metallic regime. Our theoretical \textsl{ab initio} prediction is $E = -0.0014$ Ry (see Figure~\ref{exxnew} (A)) in perfect agreement with experimental value but with an opposite sign. For a critical value of $\varepsilon_{xx}$ = 1.35 $\%$, the $K_1$ anisotropy parameter changes the sign, indicating that the magnetization direction prefers to align along [001] direction than along [100] one. The [001] direction is an easy axis of magnetization, normally observed in (Ga,Mn)As films grown on (Ga,In)As substrates that are under tensile strain \cite{PhysRevB.67.205204}.

\begin{figure}
\centering
\includegraphics[width=0.45\textwidth]{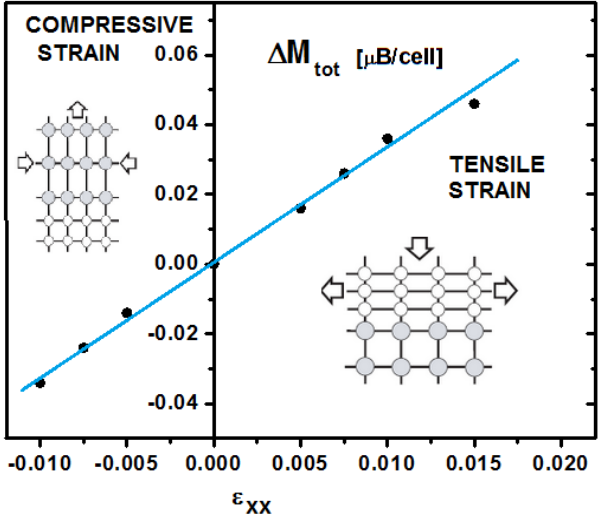}
\caption{\label{Mageexx} The influence of the biaxial strain on the value of the total magnetization vector. The difference of the total magnetization vector with respect to the magnetization of the unstrained system, as a function of the biaxial strain is presented. An increase of the interatomic distances between the atoms in-plane (001) (due to the increase of the biaxial strain), reduces the induced magnetic moments in the anionic sublattice of (Ga,Mn)As, which are the distance dependent and have an opposite sign to the Mn-moments. Therefore, the total magnetization vector increases.}
\end{figure}

A non-linear dependence of an in-plane anisotropy parameter $K_2$ as a function of biaxial strain is presented in Figure~\ref{exxnew} (B). Under the compressive strain the $K_2$ anisotropy parameter is decreasing faster than under the tensile one, for which the $K_2$ is around 80 kJ/$m^3$ when strain approaches 2$\%$. That result indicates that for a reasonable\footnote{From experimental point of view the biaxial strain $\varepsilon_{xx}$ in (Ga,Mn)As films can vary from 0$\%$ to 0.5$\%$ by changing the indium doping in the substrate \cite{PhysRevB.77.165204}.} range of biaxial strains, it is impossible to change the sign of $K_2$ parameter.

The linear dependence of the magnitude of the total magnetization vector versus the biaxial strain is presented in Figure~\ref{Mageexx}. An increase of the interatomic distances in-plane (001) between the atoms which is caused by the increase of the biaxial strain, reduces the induced magnetic moments in the anionic sublattice of (Ga,Mn)As. These induced magnetic moments are the distance dependent and have an opposite sign to the Mn-moments, hence, lower the value of total magnetization vector of the system. Now let us look at the results more precisely and determine the change of the easy and hard axis of magnetization as the function of the biaxial strain \footnote{We want to make a comment here. In the first approximation, the anisotropy parameters $K_1$ and $K_2$ indicate, respectively, the energy difference between the [001] and [100], and [110] and [1$\bar{1}$0] crystallographic directions.}. 
 
 \begin{figure}
 \centering
 \includegraphics[width=0.48\textwidth]{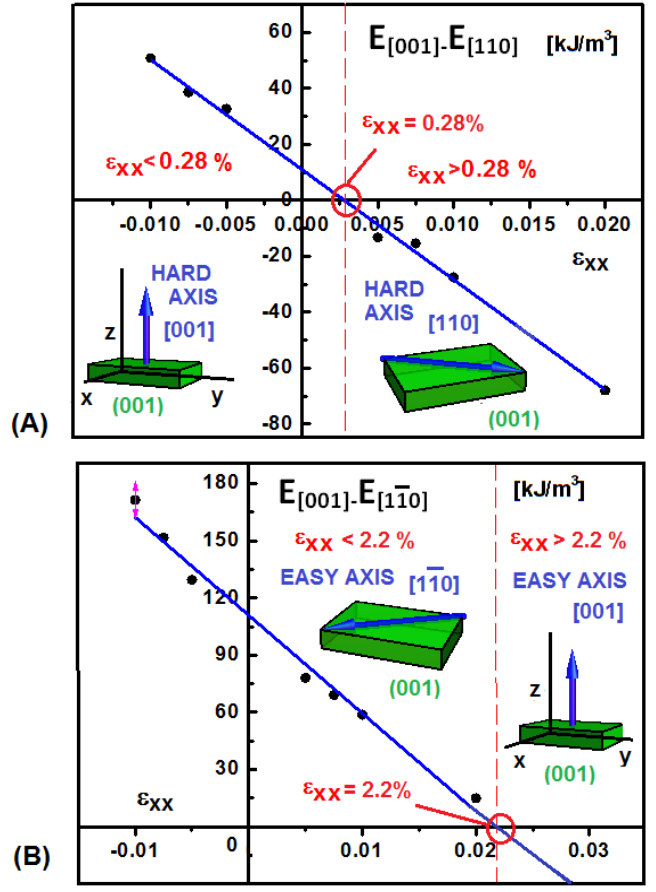}
 \caption{\label{barier} The energy barrier between two high symmetry orientations of magnetization vector,respectively for the: (A) [001] and [110], and (B) [001] and [1$\bar{1}$0] directions. The energy barrier decreases with increasing biaxial strain. (A) For the $\varepsilon_{xx}$ $<$0.28$\%$ the hard axis of the magnetization vector is [001], whereas for the $\varepsilon_{xx}$ $>$ 0.28$\%$ the [110] become a hard axis. (B) For the $\varepsilon_{xx}$ $<$ 2.2$\%$ the easy axis of the magnetization vector is [1$\bar{1}$0], whereas for the $\varepsilon_{xx}$ $>$ 2.2$\%$ the [001] becomes an easy axis. The blue arrow indicates the direction of magnetization vector in (001) plane.}
\end{figure}

\begin{table*}[ht]\footnotesize
\caption{Easy and hard axis for magnetization direction at different biaxial strain conditions; number 1 indicates the easiest axis, and the hardest axis is indexed by integer 4, [001] etc., are the high symmetry directions and $x$ values indicate that the high symmetry orientation was mentioned in the referenced paper, but weren't explicitly compared with the others' direction there. In the table, the experimental results are presented only for annealed samples.}
\label{orient}
\begin{center}
  \begin{tabular}{ | c | c | c |c|c|c|c|c| } 
\hline
 Reference&  hole concentration [$cm^{-3}$] &  Temp. [K]    &  strain $\varepsilon_{xx}$ [$\%$] & [001] & [100] & [110] &[1$\bar{1}$0] \\ 
\hline
\hline
  this work &    1.48 $10^{21}$    &    0           &    $\varepsilon_{xx}$ $<$ 0.28         &  4     &   2    &  3     &  1      \\
\hline
  this work &    1.48  $10^{21}$   &     0          &    0.28 $<$ $\varepsilon_{xx}$ $<$ 1.35 &  3     &  2     &  4     &  1      \\
\hline
  this work &    1.48  $10^{21}$   &     0          &    1.35 $<$ $\varepsilon_{xx}$ $<$ 2.2 &   2    &   3    &  4     &    1    \\
\hline
  this work &    1.48  $10^{21} $  &      0         &     $\varepsilon_{xx}$ $>$ 2.2          &   1    &  3     &   4    &   2     \\
\hline
 Exp. \textsl{Khazen at al.} \cite{PhysRevB.77.165204}   &   $10^{21} $    &  4         &     $\varepsilon_{xx}$=-0.17           &  4     &  1     &   3    &   2     \\
\hline
Exp. \textsl{Glunk at al.} \cite{PhysRevB.79.195206}   &  0.6 $10^{21} $    &  4.2       &     $\varepsilon_{xx}$=-0.22           &  4     &  1     &   x    &   x     \\
\hline
Exp. \textsl{Glunk at al.} \cite{PhysRevB.79.195206}   &  0.6 $10^{21} $    &  4.2       &     $\varepsilon_{xx}$=0.44          &  1     &  4     &   x    &   x     \\
\hline
  \end{tabular}
\end{center}
\end{table*}

The energy barriers between two high symmetry orientation of magnetization vectors versus the biaxial strain are presented in Figure~\ref{barier}. The comparison between all four high symmetry orientations is presented in Tab. \ref{orient}, from the easiest to the hardest axis of magnetization. For biaxial strain smaller than the critical value of 0.28 $\%$ (see Figure~\ref{barier}(A)), the magnetization direction [001] is a hard axis of the magnetization vector, 
and then becomes an easy axis of the magnetization vector when the strain exceeds the value of 2.2 $\%$ (see Figure~\ref{barier}(B)). Our theoretical results display the well-known experimental fact that, for high hole concentration and low temperatures, the magnetic hard axis along [001] in compressively strained layers turns into an easy axis in tensily strained layers.

This results clearly show that magnetic properties of (Ga,Mn)As films could be changed significantly by modification of the substrate on which the (Ga,Mn)As is grown. It is a well-known experimental fact that the magnitude and the sign of this anisotropy in (Ga,Mn)As can be adjusted continuously by using In$_{y}$Ga$_{1-y}$As buffer layers, where the In-concentration $y$ determines the lattice parameter of the buffer layer \cite{PhysRevB.79.195206}.

\subsection{Mn-pair with shear strain}

\begin{figure} 
\centering
\includegraphics[width=0.45\textwidth]{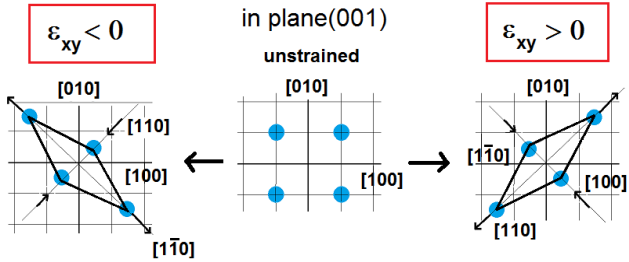}
\caption{\label{exyElong} Shear strain. At the center, a non-deformed two-dimensional square lattice is shown. The positive and negative shear strains are applied, respectively, on the right-hand side and the left-hand-side of the picture. The elongation and contraction are visible in the [110] and [1$\bar{1}$0] crystallographic
direction depending on the sign of the strain.}
\end{figure}

\begin{figure}[t] 
\centering
\includegraphics[width=0.45\textwidth]{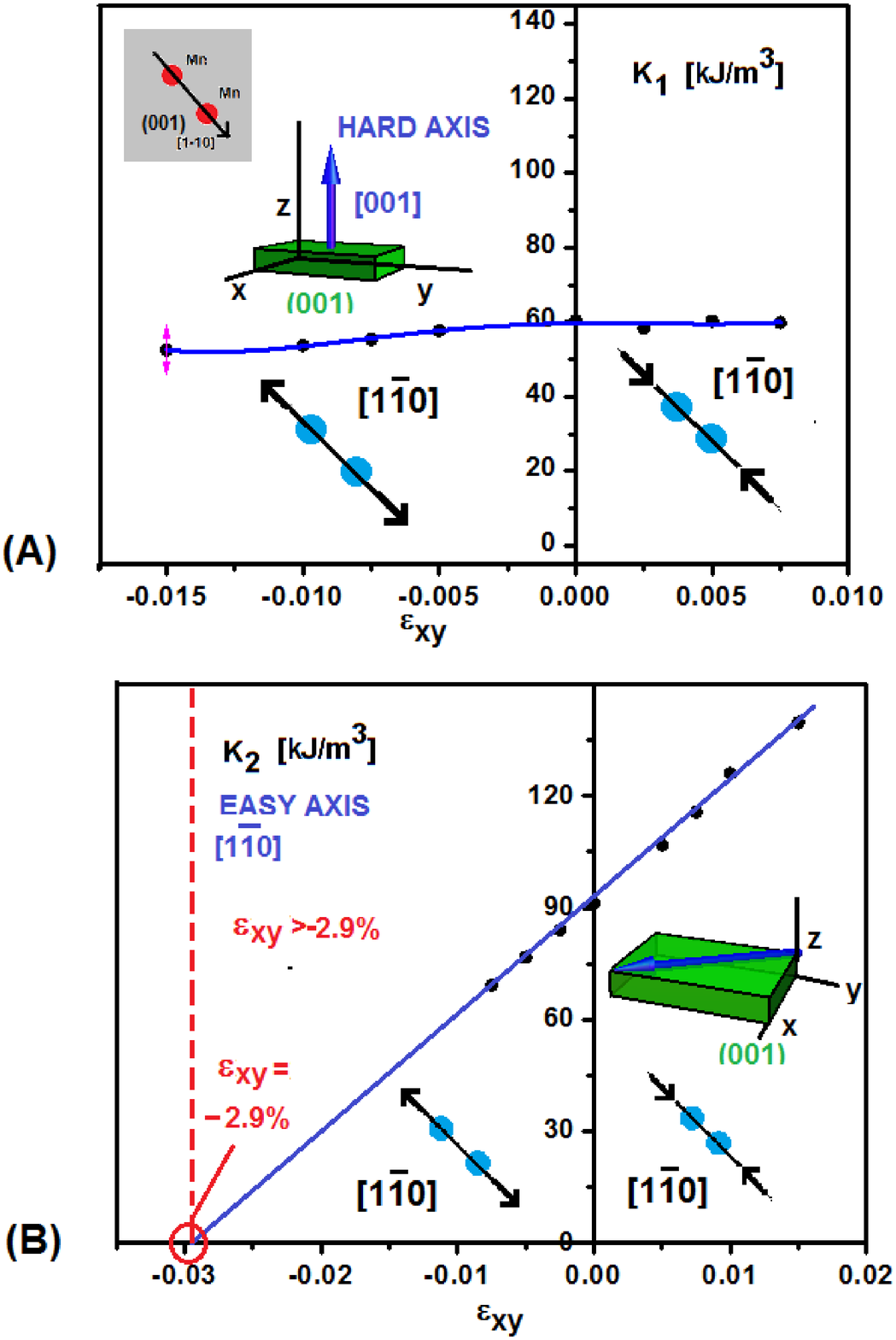}
\caption{\label{exynew1} The influence of the shear strains on the uniaxial anisotropy parameters. Dependence of the uniaxial out-of-plane(A) $K_1$ and in-plane (B) $K_2$ anisotropy parameters as a function of a shear strain $\varepsilon_{xy}$ for a Mn-concentration of x = 6.25$\%$. Linear dependency is clearly visible for $K_2$, whereas the $K_1$ is almost unchanged for a wide range of deformation's strains. The critical value indicates that the uniaxial $K_2$ is quenched. For the $\varepsilon_{xy}<-2.9\%$ the easy axis of magnetization points along [110] direction, whereas for the $\varepsilon_{xy}>-2.9\%$ along [1$\bar{1}$0] direction. The red balls on the grey area at the top of the plot illustrate the orientation of the Mn-pair in the supercell, whereas the blue balls visualize the elongation and contraction of the atoms at the [1$\bar{1}$0] direction.}
\end{figure}

Now let us turn our attention to the shear strain and its influence on a magnetic anisotropy in (Ga,Mn)As. From experimental point of view, there is no way to change the shear strain in the films so far. Therefore, the theoretical considerations are needed to find out how the specific relaxations of the atoms can influence the uniaxial magnetic anisotropy. Particularly, the theoretical calculations can give the answer how big the shear strain is needed to quench the uniaxial in-plane anisotropy parameter $K_2$. Our results show, that the out-of-plane anisotropy parameter $K_1$ is nearly independent of the shear strain for a given range of shear strains, as presented in Figure~\ref{exynew1} (A). The in-plane anisotropy parameter $K_2$ depends on shear strain linearly (see Figure~\ref{exynew1} (B)).

Note that the positive value of shear strain indicates the contraction along the [1$\bar{1}$0] direction, whereas negative its elongation along the [1$\bar{1}$0] one, which is shown schematically in Figure~\ref{exyElong}. The extrapolated critical shear strain component $\varepsilon_{xy} = -2.9$\% reflects the quenching of the uniaxial $K_2$ anisotropy parameters, which is the result of increasing average distance by about 0.12 \AA{} between two nearest neighbor atoms in the unit cell along the [1$\bar{1}$0] crystallographic direction.

\subsection{Mn-pair with hydrostatic strain}
\begin{figure} [t]
\centering
\includegraphics[width=0.42\textwidth]{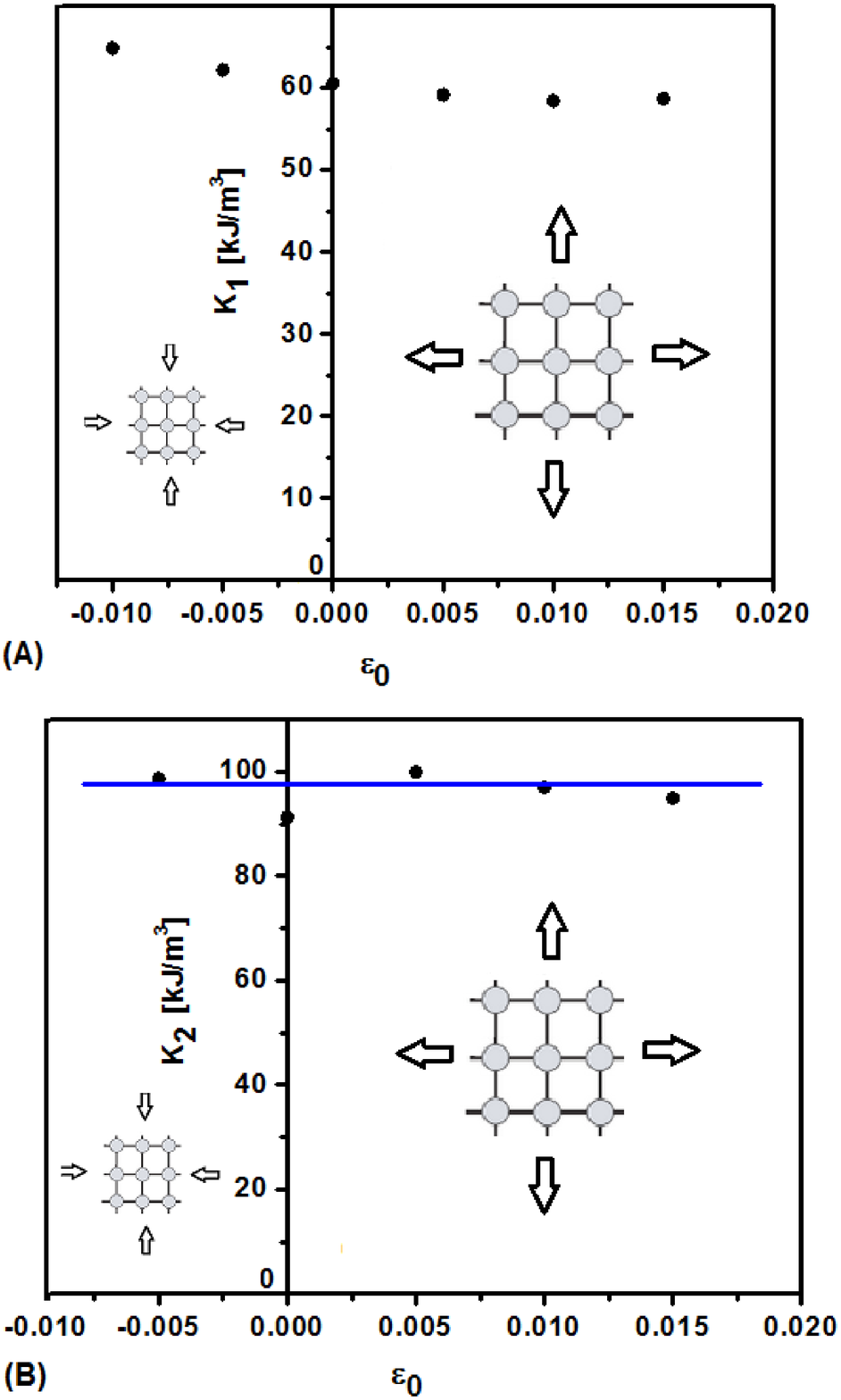}
\caption{\label{hydrostatic}The influence of the hydrostatic strain on the uniaxial anisotropy parameters. The influence of the hydrostatic strain on the uniaxial terms: (A) out-of-plane $K_1$, and (B) in-plane $K_2$ anisotropy parameters for a Mn-concentration of x = 6.25$\%$. The uniaxial terms are almost unchanged under the hydrostatic pressure. The line in the bottom part of panel (B) is only guide to the eye.}
\end{figure}

As far as we are aware, up to date, there are no systematical experimental nor theoretical \textit{ab initio} studies of the impact of hydrostatic pressure on the magnetic anisotropy energy for the dilute magnetic semiconductors. To the best of our knowledge, only in the paper \cite{Csontos.4.2005}, the authors showed that the ferromagnetic exchange interaction in (In,Mn)Sb can be controlled via the application of hydrostatic pressure. Our results show (see Figure~\ref{hydrostatic}) that under a wide range of hydrostatic pressure, the uniaxial terms $K_1$ and $K_2$ are almost unchanged, and they do not point out the way to manipulate or control the MAE via the hydrostatic pressure.

\subsection{Discussion of the results - comparison with experiments}

It is experimentally known that the magnetic anisotropy is highly sensitive to the hole concentration, temperature, and strain \cite{PhysRevB.81.155203}. Since the magnetic anisotropy sensitively depends on the individual growth conditions, care has to be taken when comparing the numerous experimental values of anisotropy parameters published by different groups.

Comparison of those data with our \textsl{ab initio} results is not so straightforward, mostly due to the presence of the unintentional defects like double donors or interstitial Mn-atoms, which are generated by LT-MBE growth conditions. In addition, the quantitative information about these intrinsic defects in real samples are not available. For example, the Mn interstitials tend to form the pairs with the substitutional Mn-atoms as-grown system, resulting in the zero net moment of the pair. Moreover, for the as-grown samples the partial concentration of interstitial Mn increases with the total Mn-concentration. We want to emphasize here that considered concentration of Mn-ions corresponds to the number of Mn-atoms substituted in the Ga-sites, whereas in the experiments, it is referred to the total Mn-doping with Mn-atoms in the interstitial sites included.

If we compare \textsl{ab initio} approach with the results obtained by the effective Hamiltonian\footnote{A group theory analysis showed that the lowering of zinc-blende symmetry due to preferable Mn incorporation onto the surface, leads to the appearance of two additional terms in
 the \textsl{k-p} Hamiltonian, being of the form of effective biaxial $\varepsilon_{xx}$ and shear $\varepsilon_{xy}$ strains, respectively.} \cite{PhysRevLett.108.237203}, 
 for the Mn-concentration of $x = 6.25 \%$ the perturbation approach gives the values of
 $K_2 = -2.7$ [kJ/$m^3$] and $K_1 = -5.1$ [kJ/$m^3$]. These values are one order of magnitude lower and have opposite signs than the results obtained from \textsl{ab initio} approach. 
The \textsl{ab initio} results take into account the relaxation of the atoms in the supercell and do not take into account the external parameters, whereas the perturbation approach does not include the microscopic atomistic structure of the system.

Now let us turn our attention to compare of our results with the relative strength of the experimentally obtained anisotropy components. Typical values of the uniaxial anisotropy parameters experimentally measured are in the range of few $kJ/m^3$. Our values for the unstrained supercell for temperature $T=0$ K, and concentration of the Mn-atoms $6.25\%$ are one order of magnitude higher than experimentally obtained. The reason for this discrepancy seems to be the assumption that all of the Mn-pairs reside entirely along [1$\bar{1}$0] crystallographic direction, however in a real system, there are also presumably the pairs along the [110] direction and a smaller number of pairs along different directions. In real samples, there are always some double donors. This lowers the concentration of holes, and hence it decreases the magnetic anisotropy. 

We would like to note that also in extremely thin (Ga,Mn)As layers contributions from interfacial and surface anisotropies (for them, the magnetic uniaxial anisotropy is allowed even for a random distribution of magnetic ions), may gradually come into play.

It is worth to mention that our results show that one does not need any external strain, trigonal or biaxial, to explain observed in- and/or out-of-plane uniaxial magnetic anisotropy terms, as it was also mentioned in the paper \cite{PhysRevLett.108.237203}.

\section{Summary}

We have studied the influence of the different external strains on the magnetic anisotropy energy for a 6.25 \% concentration of the Mn-ions at the nearest neighbor position in the supercell.
Our results show that there exists the linear trend of the out-of-plane anisotropy parameter $K_1$ versus the biaxial strain, which is consistent with the previous theoretical approach and experimental findings. We have demonstrated that there exists a critical biaxial strain for which the magnetization vector can switch its direction. Moreover, our results have displayed the well-known experimental fact that, for high hole concentration and low temperatures, the magnetic hard axis along [001] in compressively strained layers turns into an easy axis in tensily strained layers. Furthermore, our predicted biaxial critical values are much higher than for experimental results, which is directly related with our assumption that the pairs are placed at a given position at the crystal.

In case of the shear strain, we have obtained the critical shear value for which  the quenching of the uniaxial magnetic anisotropy parameter $K_2$ exists. It reflects the increase of the distance between the Mn-ions at the [1$\bar{1}$0] crystallographic direction, for which the uniaxial magnetic anisotropy in-plane vanishes. Moreover, we have predicted that it is not possible to modify considerably the uniaxial magnetic anisotropy parameters, exposing (Ga,Mn)As to hydrostatic pressure of a magnitude reasonable from experimental point of view.

\section*{Acknowledgements}

This work was supported by the European Research Council through the FunDMS Advanced Grant within the "Ideas" Seventh Framework Programme of the EC and InTechFun (POIG 01.03.01-00-159/08) of EC. We made use of computing facilities of PL-Grid Polish Infrustructure for Supporting Computational Science in the European Research Space, and acknowledge the access to the computing facilities of the Interdisciplinary Centre of Modeling, University of Warsaw. The support of the National Research Council (NCN) through the grant HARMONIA DEC-2013/10/M/ST3/00793 is gratefully acknowledged.

\section*{References}

\bibliographystyle{elsarticle-num}
\bibliography{Magnetism}

\end{document}